\documentclass[reprint,superscriptaddress,showpacs,nofootinbib,amsmath,amssymb]{revtex4-1}
\usepackage{comment}
\usepackage{amssymb}
\usepackage{color}
\usepackage{rotating}
\usepackage{amsmath}
\usepackage{ulem}
\usepackage{overpic}
\usepackage{graphicx}
\usepackage{dcolumn}
\usepackage{bm}
\usepackage{epstopdf}
\usepackage{chngpage}
\usepackage{multirow}
\usepackage{slashed}
\usepackage{indentfirst}
\usepackage{booktabs}
\usepackage{amssymb,amsfonts}
\usepackage{amsmath}
\usepackage{color}
\usepackage[colorlinks,citecolor=blue,anchorcolor=blue,linkcolor=blue,hypertex,breaklinks=true]{hyperref}

\begin{document}
\title{Improved description of nuclear charge radii: Global trends beyond $N=28$ shell closure}
\author{Rong An}
\affiliation{School of Physics, Ningxia University, Yinchuan 750021, China}
\affiliation{Key Laboratory of Beam Technology of Ministry of Education, College of Nuclear Science and Technology, Beijing Normal University, Beijing 100875, China}

\author{Xiang Jiang}
\affiliation{College of Physics and Optoelectronic Engineering, Shenzhen University, Shenzhen 518060, China}

\author{Na Tang}
\affiliation{School of Physics, Ningxia University, Yinchuan 750021, China}

\author{Li-Gang Cao}
\email[Corresponding author: ]{caolg@bnu.edu.cn}
\affiliation{Key Laboratory of Beam Technology of Ministry of Education, College of Nuclear Science and Technology, Beijing Normal University, Beijing 100875, China}
\affiliation{Key Laboratory of Beam Technology of Ministry of Education, Institute of Radiation Technology, Beijing Academy of Science and Technology, Beijing 100875, China}

\author{Feng-Shou Zhang}
\email[Corresponding author: ]{fszhang@bnu.edu.cn}
\affiliation{Key Laboratory of Beam Technology of Ministry of Education, College of Nuclear Science and Technology, Beijing Normal University, Beijing 100875, China}
\affiliation{Key Laboratory of Beam Technology of Ministry of Education, Institute of Radiation Technology, Beijing Academy of Science and Technology, Beijing 100875, China}
\affiliation{Center of Theoretical Nuclear Physics, National Laboratory of Heavy Ion Accelerator of Lanzhou, Lanzhou 730000, China}

\begin{abstract}
  Charge radii measured with high accuracy provide a stringent benchmark for characterizing nuclear structure phenomena.
  In this work, the systematic evolution of charge radii for nuclei with $Z=19$-$29$ is investigated through relativistic mean field theory with effective forces NL3, PK1, and NL3$^{*}$.
  The neutron-proton ($np$) correlation around Fermi surface originated from the unpaired neutron and proton has been taken into account tentatively in order to reduce the overestimated odd-even staggering of charge radii.
  This improved method can give an available description of charge radii across $N=28$ shell closure.
  A remarkable observation is that the charge radii beyond $N=28$ shell closure follow the similarly steep increasing trend, namely irrespective of the number of protons in the nucleus.
  Especially, the latest results of charge radii for nickel and copper isotopes can be reproduced remarkably well.
  Along $N=28$ isotonic chain, the sudden increase of charge radii is weakened across $Z=20$, but presented evidently across $Z=28$ closed shell.
  The abrupt changes of charge radii across $Z=22$ are also shown along $N=32$ and $34$ isotones, but the latter with a less slope.
  This seems to provide a sensitive indicator to identify the new magicity of a nucleus with universal trend of charge radii.
\end{abstract}



\maketitle
\section{INTRODUCTION}\label{sec0}
Nuclear charge radius encodes the information about the charge density distribution in a self-bound many-nucleon system.
Reliable description of nuclear charge radius can provide access to recognize various pronounced structure phenomena, such as halo structures~\cite{PhysRevLett.101.252502,PhysRevLett.102.062503}, shape staggering~\cite{Marsh:2018wxs,PhysRevC.95.044324,PhysRevC.99.044306,PhysRevLett.127.192501}, the onset of shape deformation~\cite{LALAZISSIS199635,PhysRevLett.54.1991,PhysRevLett.117.172502}, and the emergence of nuclear magicity~\cite{KREIM201497,Gorges2019,PhysRevC.102.051303,BAGCHI2019251,PhysRevLett.129.142502}, etc.
Meanwhile, charge radii difference of mirror-pair nuclei can provide a probe to constrain the isovector components of equation of state (EoS) in asymmetric nuclear matter~\cite{PhysRevC.88.011301,PhysRevLett.119.122502,PhysRevC.97.014314,PhysRevResearch.2.022035,XU2022137333,
PhysRevLett.127.182503,PhysRevLett.130.032501,PhysRevC.107.034319,nuclscitech34.119,PhysRevC.108.015802,Konig:2023rwe}.
With the improved measurements, much more data of charge radii of exotic nuclei far away from the $\beta$-stability line have been compiled~\cite{ANGELI201369,LI2021101440,Goodacre2021,YANG2023104005}.
The distinguished aspects of nuclear charge radii, involving the shell closures and odd-even staggering (OES) effects, are generally observed throughout the entire nuclide chart~\cite{CAMPBELL2016127,GarciaRuiz:2019cog}.

The discontinuity changes in charge radii are naturally observed across the traditional neutron magic numbers $N=28$, $50$, $82$, and $126$~\cite{PhysRevC.86.034329,Gorges2019,SHARMA19939,PhysRevLett.110.032503,PhysRevLett.121.102501,deGroote}.
Many methods pertaining to the effect of shell closure on nuclear charge radii are discussed~\cite{PhysRevC.88.011301,PhysRevC.100.044310,PhysRevC.104.064313}.
As demonstrated in Ref.~\cite{PhysRevLett.74.3744}, this shell closure effect results from the rather small isospin dependence of the spin-orbit term.
Especially striking among these fully filled shells has been found is the intermediate mass region nuclei featuring the $N=28$ neutron-closure shell.
The inverted parabolic-like shape between $N=20$ and $N=28$ and odd-even staggering (OES) in charge radii are observed remarkably along Ca isotopes.
The same scenario can also be found in K isotopes, but with weakened amplitudes~\cite{ANGELI201369}.
Across the $N=28$ shell closure, the experimentally observed strong increases in charge radii are naturally manifested for K~\cite{PhysRevC.100.034304,Koszorus2020mgn} and Ca~\cite{Ruiz2016} isotopes,
and the odd-even oscillation behaviors have been weakened as well.
Recently, charge radii of $^{55,56,58-68,70}\mathrm{Ni}$ isotopes have been measured by collinear laser spectroscopy method~\cite{nickle,PhysRevLett.129.132501}.
It suggests that the trend of charge radii across the $N=28$ shell closure is similar to Ca isotopes.
Between neutron numbers $N=28$ and $N=40$, the nuclear charge radii exhibit a universal pattern that is independent of atomic number~\cite{PhysRevC.105.L021303}.
These local variations pose great challenges to our knowledge of understanding nuclear force.

The undertaken efforts have been devoted to describing the global trend of charge radii in nuclear chart.
The sophisticated algebraic expression involving the Casten factor $P$ can also reproduce the shell effects~\cite{PhysRevLett.58.658,Angeli_1991}.
Subsequently, a five-parameter formula has been proposed by introducing the OES effects~\cite{Sheng:2015poa},
in which neutron-proton ($np$) interaction originating from the valence proton and neutron contributes to the local variation of nuclear size.
This indicates that the $np$ correlation plays an essential role in accessing the fine structure of nuclear size.
Miller, $et~al.$ point out that short-range neutron-proton tensor interactions cause the protons to move closer to the outside neutrons of a nucleus, thereby increasing the charge radius~\cite{Miller:2018mfb}.
This emphasizes that $np$ correlation results in the changes of the calculated root-mean-square (rms) charge radius.
This short-range correlations (SRCs) imply a decrease in the occupation of the states below the Fermi level and a partial occupation in the states above it~\cite{PhysRevC.101.065202}.
Recent study further verifies that the nuclear SRCs induce the high-momentum fluctuations in the nuclear medium, and the role of $np$-correlation has a non-negligible influence on nuclear size~\cite{Ryckebusch2021}.

As demonstrated in Ref.~\cite{Zawischa:1985qds}, there is an attractive interaction between proton and neutron pairs, the origin of OES behaviors in nuclear charge radii is due to the four-particle correlations (or being $\alpha$ cluster).
The relativistic density functionals cannot describe the fine structure of charge radii along calcium isotopic chain well~\cite{Geng:2003pk,BLAUM200830,XIA20181,Zhang2022ADNDT}.
To reproduce the OES and shell effect of charge radii along calcium isotopes, a modified relativistic mean field plus Bardeen-Cooper-Schrieff (BCS) equation ansatz, namely RMF(BCS)*, has been proposed~\cite{An:2020qgp}.
In which the semi-microscopic correction originating from the neutron and proton pairs condensation is introduced by solving the state-dependent BCS equations.
The calculated results are in good agreement with the data and compatible with the sophisticated Fayans energy density functional (EDF) approach where the surface pairing dominates a role in reproducing the local variation of charge radii~\cite{Reinhard2017,Miller2019}.
However, for odd-$Z$ isotopic chain, the OES effects in charge radii are overestimated by RMF(BCS)* approach~\cite{an2021charge}.
This may result from the fact that the correlation between the unpaired neutron and proton due to the violation of time-reversal symmetry is ignored.
The same scenarios can also be encountered in Fayans EDF model where the OES effects in charge radii are overestimated for K~\cite{Koszorus2020mgn} and Cu~\cite{deGroote} isotopes as well.
Obviously, the influence coming from the unpaired nucleons is still excluded out the pairing spaces due to the blocking approximation.

So far, the existing theoretical models can hardly give a unified description of charge radii around $N=28$ mass region.
Hence the required model should be able to describe simultaneously the shell effect and OES in nuclear charge radii.
In this work, the charge radii are calculated along $Z=19$-$29$ isotopic chains that include the nuclei featuring the $N=28$ magic shell and the new magicity of $N=32$ and $34$.
In order to reduce the overestimated OES behaviors of charge radii along odd-$Z$ isotopic chains, the $np$ correlation originating from the simultaneously unpaired neutron and proton is taken into account in the improved charge radii formula.
Generally, the gradually shrunken trend of nuclear charge radii is observed around the traditional neutron magic numbers along a long isotopic chain~\cite{ANGELI201369,LI2021101440}.
One should also mention the great interest for the properties of the neutron number $N=32$ and $34$ isotopes that have been proposed as new-magicity nuclei in certain isotopic chains.
As demonstrated in Ref.~\cite{PhysRevC.98.024310}, the new magicity of $N=32$ is vanished beyond $Z=22$. Meanwhile, the sub-shell closure effect of $N=34$ is weakened across $Z=20$ proton number~\cite{PhysRevC.82.014311,PhysRevC.96.064310,PhysRevC.101.052801}.
The sensitive indicators of new-magicity associating with the evolution of nuclear charge radii are also paid more attention along $N=32$ and $34$ isotonic chains in our calculations.

The structure of the paper is the following.
In Sec.~\ref{sec1}, the theoretical framework is briefly presented.
In Sec.~\ref{sec2}, the numerical results and discussion are provided.
Finally, a summary is given in Sec.~\ref{sec3}.

\section{Theoretical framework}\label{sec1}
The relativistic mean-field (RMF) theory with different parametric versions has made considerable success in describing various phenomena in nuclear physics~\cite{PhysRevLett.77.3963,Vretenar:2005zz,Liang:2014dma,jie2016relativistic,Cao:2003yn,PhysRevC.67.034312,PhysRevC.69.054303,PhysRevC.68.034323,MENG2006470,PhysRevC.82.011301,PhysRevC.85.024312,
PhysRevC.92.024324,NIKSIC20141808,SUN2018530,PhysRevC.102.024314,Zhang:2021ize,Rong:2023tdl}.
In this work, the standard non-linear self-coupling Lagrangian density is employed~\cite{Ring:1997tc}.
The Dirac equations with effective fields $S(\mathbf{r})$ and $V(\mathbf{r})$ are derived through variational principle.
To obtain the ground-state properties of finite nuclei, the quadrupole deformation parameter $\beta_{20}$ becomes constrained in the self-consistent iterative process.
Therefore, the Hamiltonian formalism can be recalled as follows:
\begin{eqnarray}
\mathcal{H}=-i\alpha\nabla+V(\mathbf{r})+\beta{[M+S(\mathbf{r})]}-\lambda\langle{\mathbf{Q}}\rangle,
\end{eqnarray}
where $M$ is the mass of nucleon, $\lambda$ is the spring constant, and $\mathbf{Q}$ is the intrinsic quadrupole moment.
Those values of $\beta_{20}$ are changed from $-0.50$ to $0.50$ with the interval range of $0.01$.

Recent study shows that it is essential to include the staggering behaviors of charge radii in validation protocol~\cite{dong2021novel,DONG2023137726}.
As mentioned above, a new ansatz has been proposed for describing the OES and shell closure effects in the charge radii of calcium isotopes~\cite{An:2020qgp}.
This modified formula can also reproduce the global trend of charge radii along various even-$Z$ isotopic chains~\cite{PhysRevC.105.014325}.
Although the shell effects and OES in nuclear charge radii are remarkably described well, the validation protocol is just considered by the chosen nuclei $^{44}$Ca and $^{126}$Sn~\cite{An:2020qgp}.
Besides, the $np$ correlation originating from the simultaneously unpaired proton and neutron is excluded; this leads to the overestimated OES of charge radii for odd-$Z$ isotopic chains~\cite{an2021charge,An064101}.
Therefore, the improved version should be proposed, especially the $np$ correlation coming from the unpaired nucleons must be considered.
To account for the implications of the observed odd-even oscillations of charge radii along odd-$Z$ isotopic chains, the further modified mean-square charge radii formula is proposed as follows (in units of fm$^2$),
\begin{eqnarray}\label{cp2}
R_{\mathrm{ch}}^{2}=\langle{r_{\mathrm{p}}^{2}}\rangle+0.7056+\frac{a_{0}}{\sqrt{A}}\Delta{\mathcal{D}}~\mathrm{fm^{2}}+\frac{\delta}{\sqrt{A}}~\mathrm{fm^{2}}.
\end{eqnarray}

The first term represents the charge distribution of point-like protons and the second term is due to the finite size of protons~\cite{Gambhir:1989mp}.
Here, the quantity of the proton radius takes the values about $0.84$ fm~\cite{RevModPhys.93.025010,PhysRevLett.128.052002}.
As shown in Eq.~(\ref{cp2}), the quantity of $\mathcal{D}$ is associated with the Cooper pairs condensation~\cite{PhysRevC.76.011302}.
Furthermore, this quantity can be used to measure the Fermi surface diffuseness encoded by various eigenfunctions~\cite{MANG1965353}.
Its value is calculated self-consistently by solving the state-dependent BCS equations~\cite{Geng:2003pk,RongAn:114101}.
The difference of $\mathcal{D}$ coming from neutrons and protons has been attributed to the $np$ correlation around Fermi surface~\cite{An:2020qgp,PhysRevC.105.014325}.

It is mentioned that the amplitude of the OES of charge radii in potassium isotopes is obviously lower than the case of calcium isotopes due to the last unpaired proton~\cite{ANGELI201369,LI2021101440}. For the theoretical studies, the developed RMF(BCS)* approach and Fayans EDF model can describe the OES of charge radii along even-$Z$ isotopic chains well~\cite{An:2020qgp,Reinhard2017}. However, along odd-$Z$ isotopic chains, the overestimated OES in charge radii is obtained by both of the RMF(BCS)* approach and Fayans EDF model~\cite{deGroote,an2021charge,Koszorus2020mgn}. The correlation between the simultaneously unpaired proton and neutron may melt this tension.
The last term in this expression just represents the $np$ correlation deriving from the simultaneously unpaired neutron and proton.
This means the quantity of $\delta$ equals to zero for even-even, odd-even, and even-odd nuclei.
The values of $a_{0}=0.561$ and $\delta=0.355$ are calibrated by fitting the odd-even oscillation and the inverted parabolic-like shape of charge radii along potassium and calcium isotopes under effective force NL3.
And then the extrapolated calculations are performed for $Z=21$-$29$ isotopic chains.
The PK1 set gives a more reasonable baryonic saturation density and a reasonable description of spin-orbit splittings and single-particle energies~\cite{PhysRevC.69.034319}.
In addition, NL3$^{*}$ parameter set improves the description of the ground-state properties of many nuclei and simultaneously provides an excellent description of the collective excited states properties in finite nuclei~\cite{Lalazissis:2009zz}.
Thus these two effective forces are also employed to access the global evolution of charge radii along $Z=19$-$29$ isotopic chains.
It should be particularly interesting for quantitatively understanding the reduced OES of charge radii along odd-$Z$ isotopic chains and the trend of changes of nuclear charge radii across the $N=28$ shell closure.

\section{Results and Discussions}\label{sec2}
In this work, we focus on the systematic behaviors of charge radii along $Z=19$-$29$ isotopic chains.
The bulk properties of finite nuclei are calculated by relativistic mean field theory with NL3~\cite{PhysRevC.55.540}, PK1~\cite{PhysRevC.69.034319}, and NL3$^{*}$~\cite{Lalazissis:2009zz} effective forces.
The pairing correlations are treated by the state-dependent Bardeen-Cooper-Schrieffer method, which can capture the ground-state properties of finite nuclei well~\cite{Geng:2003pk}.
The pairing strength is determined through the empirical odd-even mass staggering of calcium isotopes~\cite{Bender:2000xk}.
To reflect the universality of our results, the pairing strength is chosen to $V_{0}=350$ MeV fm$^{3}$ for NL3 set, but $V_{0}=380$ MeV fm$^{3}$ for both PK1 and NL3$^{*}$ interaction forces, respectively.
Based on the fact that the nuclei with featuring new magicity have been observed at the neutron numbers $N=32$ and $34$.
It is instructive to investigate the universal behaviors of charge radii along $N=28$, $32$ and $34$ isotones.

\subsection{Shell quenching of nuclear charge radii at $N=28$ }
\begin{figure}[htbp!]
\includegraphics[scale=0.4]{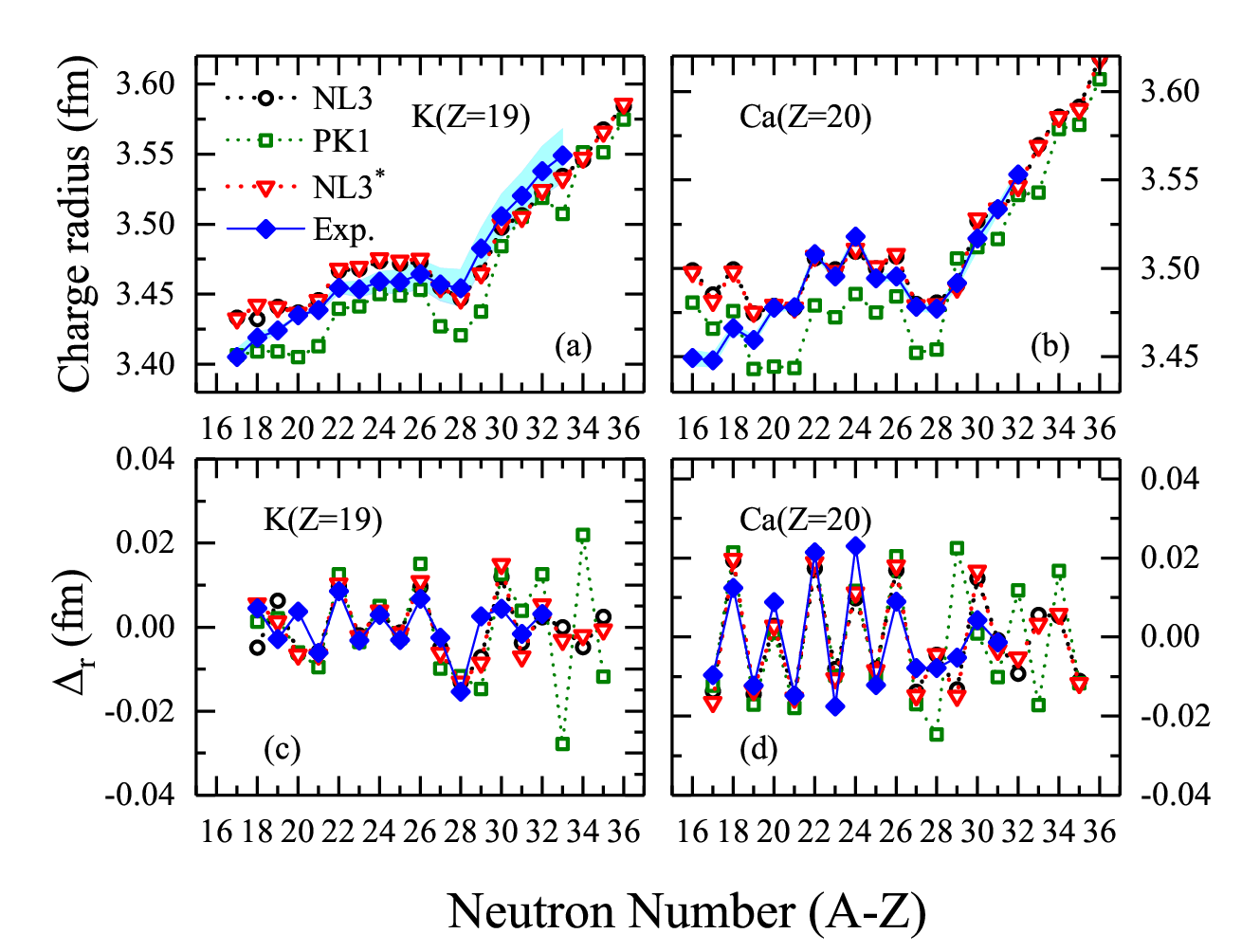}
    \caption{Evolution in rms charge radii versus neutron number for (a) potassium and (b) calcium isotopes. The effective forces NL3 (open circle), PK1 (open square), and NL3$^{*}$ (open triangle) are employed in relativistic mean-field framework. The corresponding odd-even staggerings are also depicted in (c) and (d). Experimental data are taken from Refs.~\cite{ANGELI201369,LI2021101440,Koszorus2020mgn}, which are represented by solid diamond with systematic error bands (light blue). } \label{fig1}
\end{figure}

The inverted parabolic-like behavior describing the charge radii of the calcium isotopes between $N = 20$ and $28$ as well as the amplitude of OES are much more pronounced for the neighboring odd-$Z$ element potassium~\cite{TOUCHARD1982169}.
The inverted parabolic-like curves are symmetric to the $1f_{7/2}$ mid-shell neutron number $N = 24$ for potassium and calcium isotopic chains~\cite{ANGELI201369}.
In Fig.~\ref{fig1}, the rms charge radii of potassium and calcium isotopes as a function of neutron number are plotted by the relativistic mean-field (RMF) model with effective forces NL3, PK1, and NL3$^{*}$.
One can find that the inverted parabolic-like shapes of potassium and calcium isotopes are compared against well by NL3 and NL3$^{*}$ parametrization sets, while the quantitative level of charge radii is systematically underestimated by PK1 force.
Beyond $N=28$ shell closure, the sharp increases are identified by these three forces.
In Ref.~\cite{Borzov2022}, the universal increase of the charge radii after crossing the neutron shell at $N=28$ is attributed to the quasiparticle-phonon coupling.

The characteristic aspects in charge radii are the odd-even oscillation behaviors along calcium isotopic.
Reference~\cite{TALMI1984189} demonstrates that the main contribution to the odd-even effect is ascribed to quadrupole and higher even-order core polarization,
in which the strong and attractive interaction between the valence neutrons and core protons has been considered.
These peculiarities have been ascribed to the changes of dynamic deformation for different orders~\cite{BARRANCO198590}.
Qualitatively the OES of radii as a function of $N$ is associated with the reduction of core polarization due to unpaired nucleons~\cite{REEHAL1971385}.
This mechanism also provides a straightforward explanation for the odd-even effect as a function of $Z$.
Besides, high-order radial moment offers key information on the nuclear charge radius~\cite{PhysRevC.101.021301}.

As shown in Ref.~\cite{Koszorus2020mgn}, the amplitudes of OES in charge radii for potassium isotopes are overestimated by Fayans EDF.
Fayans EDF model points out that surface pairing interactions cause the origin of OES in nuclear charge radii~\cite{Reinhard2017}.
However, the time-reversal states are excluded in tackling pairing correlations, especially for simultaneously unpaired neutron and proton in potassium isotopes.
The same scenario is encountered in Ref.~\cite{an2021charge}, where the states that violate time-reversal symmetry are tackled by the blocking approximation.
This leads to the most overestimated amplitudes of OES of charge radii in potassium isotopes as well.
In Ref.~\cite{An:2020qgp}, $np$ correlation originating from the difference of Cooper pair condensation is introduced by solving the state-dependent BCS equations.
The modified root-mean-square charge radii formula can describe well the OES and shell effects in charge radii along even-$Z$ isotopic chains.
But for odd-$Z$ cases, the OES is overestimated due to the absence of $np$ correlation coming from the simultaneously unpaired neutron and proton~\cite{an2021charge}.
As shown in Fig~\ref{fig1}, the modified formula can reproduce the OES of charge radii between $N=20$ and $28$ along potassium and calcium isotopes.

In Fig.~\ref{fig1}~(a), charge radii of $^{36-38}$K isotopes are slightly overestimated by the NL3 and NL3$^{*}$ forces.
By contrast, the calculated results for $^{50}$K are slightly underestimated by these two forces.
The experimental mass measurement suggests that $^{51}$K has a relative large shell gap with respect to the neighboring counterparts~\cite{PhysRevLett.114.202501}.
Of high interest is also the experimental determination of the charge radii across the neutron number $N=32$ with emerging the new magicity.
A signature of the magic character can be reflected from the sudden increase in charge radii.
Charge radii of exotic isotope $^{52}$K has been performed by the collinear resonance ionization spectroscopy method~\cite{Koszorus2020mgn}.
This latest measurement suggests that no rapid increase in charge radii can be found across $^{51}$K.
Beyond $^{50}$K, these three effective forces give almost similar trends and show a smooth increase from $^{51}$K to $^{52}$K.

In Fig.~\ref{fig1}~(b), charge radii of $^{37-39}$Ca isotopes obtained by PK1 force are compared well against NL3 and NL3$^{*}$ forces.
NL3$^{*}$ parameter set gives almost comparable results with respect to NL3 force.
With the increasing neutron number occupation at the $p_{3/2}$ orbital, the unexpected linear increase of the charge radius occurs~\cite{PhysRevLett.129.262501}.
This can also be used to understand the emergency of the abrupt increase of charge radii across $N=28$ shell closure as shown in Fig.~\ref{fig1}.
The neutron number $N=32$ is identified as a new magic number in calcium region~\cite{RongAn:114101,nature498,PhysRevLett.109.032506}.
Therefore, the implication of this experimental value for the charge radius of $^{53}$Ca will be urgently expected.
As encountered in potassium isotopes, the charge radii beyond $N=34$ show same steep increasing trend with these effective forces.

In order to inspect these local variations of charge radii along potassium and calcium isotopes, the three-point formula is recalled as follows~\cite{Reinhard2017,An:2020qgp},
\begin{small}
\begin{eqnarray}\label{oef1}
\Delta_{r}(N,Z)=\frac{1}{2}[2R(N,Z)-R(N-1,Z)-R(N+1,Z)],
\end{eqnarray}
\end{small}
where $R(N,Z)$ is rms charge radius for a nucleus with neutron number $N$ and proton number $Z$.
In Fig.~\ref{fig1}~1(c) and 1(d), the OES of charge radii in potassium and calcium isotopes is depicted.
For potassium isotopes, the OES in charge radii can be reproduced well across $N=20$ shell closure.
Extending to neutron-deficient regions, the obtained OES amplitudes for these three effective forces cannot reproduce the experimental data well.
This can be due to the fact that the charge radii in neutron-deficient region are overestimated by NL3 and NL3$^{*}$ sets and  underestimated by PK1 force.
For neutron-rich region, the PK1 force gives the enlarged OES of charge radii with respect to NL3 and NL3$^{*}$ sets.
Along calcium isotopes, the OES behaviors of charge radii for $^{37-47}$Ca isotopes can be reproduced well by these forces.
Across the $N=28$ filled shell, PK1 set gives the inverse OES amplitudes in charge radii with respect to NL3 and NL3$^{*}$ sets.
Besides, it can be found that the OES behavior is weakened at the neutron number $N=28$.
This is in accord with Ref.~\cite{PhysRevC.105.014325} where the weakened OES in charge radii can be regarded as signature of the emergence of neutron magic numbers.

\begin{figure}[htbp]
\includegraphics[scale=0.32]{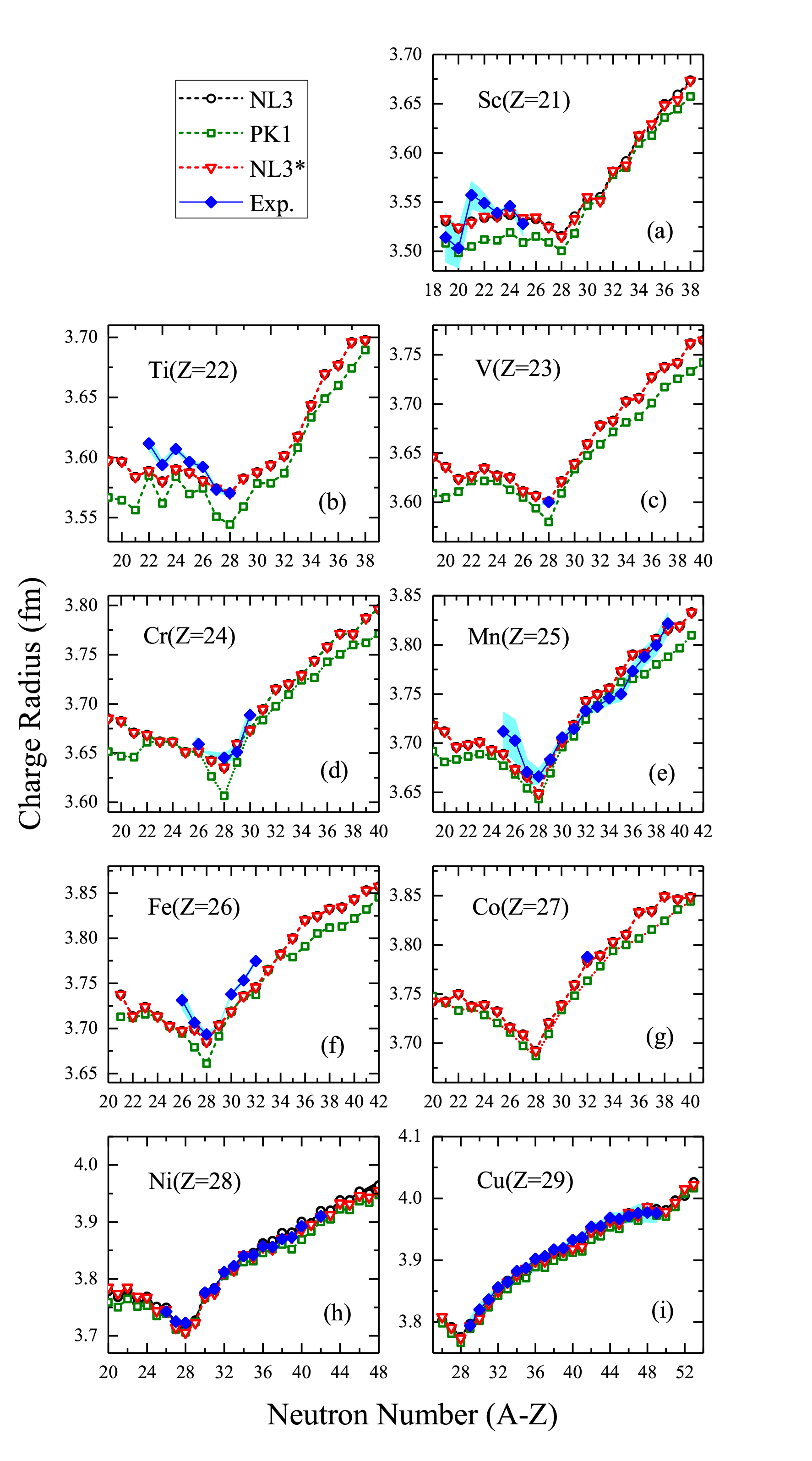}
\caption{Same as Fig.~\ref{fig1}, but for (a) scandium, (b)titanium, (c) vanadium, (d) chromium, (e) manganese, (f) iron, (g) cobalt, (h) nickel, and (i) copper isotopes. Experimental data are taken from Refs.~\cite{ANGELI201369,LI2021101440}, while the data of Sc isotopes are taken from Ref.~\cite{PhysRevLett.131.102501}, the value of $^{54}$Ni is extracted from Ref.~\cite{PhysRevLett.127.182503} and $^{58-68,70}$Ni are extracted from Ref.~\cite{nickle}.} \label{fig2}
\end{figure}
To facilitate the instructive understanding of charge radii on the isotopic chains of neighboring elements, the systematic evolution of charge radii along $Z=21$-$29$ isotopic chains is plotted in Fig.~\ref{fig2}.
From this figure, one can find that the shell effect at $N=28$ is evidently observed.
Across $N=28$, the abrupt changes are universally observed in this region, while the slope of change is lower for scandium ($Z=21$) and titanium ($Z=22$).
In Fig.~\ref{fig2} (a), these three forces give the inverted parabolic-like behaviors in charge radii between $N=20$ and $N=28$,
while PK1 parameter set slightly underestimates the results.
The latest study suggests that the charge radii of $^{41}$Sc isotope is surprisingly shrunk~\cite{PhysRevLett.131.102501}, then the abrupt increase is occurred beyond the neutron number $N=20$.
In our calculations, the PK1 force can reproduce the charge radii of $^{40,41}$Sc isotopes, but the values of $^{42-46}$Sc isotopes are exaggeratedly underestimated. By contrast, these phenomena are mostly met by the NL3 and NL3$^{*}$ forces.
The NL3 and NL3$^{*}$ forces can cover the experimental uncertainties of $^{40,41}$Sc, but the values of $^{42,43}$Sc are underestimated. It is notably mentioned that our calculations can give the trend of kink, but the amplitude is heavily reduced.
Besides, the last unpaired nucleon is just tackled by the blocking approximation, namely occupies the last single-particle level.
This may also result in the underestimated results.
The same scenarios are also encountered for titanium isotopes.
As shown in Ref.~\cite{Gangrsky_2004}, the data of the charge radii on the titanium isotopes show a continuous decrease from $N = 22$ to $28$.
The calculated results for $^{44-46}$Ti are slightly deviated from experimental data.
But the trend of changes of charge radii are not changed, especially for $^{44-46}$Ti where the odd-even oscillation behaviors can be reproduced well.

For V, Cr, and Co isotopic chains, the experimental data can be reproduced well.
However, more experimental data are urgently encouraged to support our results.
In these regions the overall slope increases gradually with the increasing neutron numbers.
Also the shell closure effect in charge radii is clearly pronounced  at $N = 28$ due to the rather small isospin dependence of spin-orbit interactions~\cite{PhysRevLett.74.3744}.
Along manganese isotopes, results obtained by NL3, PK1, and NL3$^{*}$ forces can reproduce the experimental data across $N=28$ shell closure.
However, there exists a slight deviation between experimental data and results obtained by PK1 force.
For $^{50,51}$Mn and $^{52}$Fe, the calculated results are all reduced.
Meanwhile, the data of $^{56-58}$Fe isotopes are also slightly underestimated.

Recent studies on nuclear size are performed along nickel isotopic chain~\cite{PhysRevLett.127.182503,nickle}.
In Fig.~\ref{fig2}~(h), one can find that NL3, PK1, and NL3$^{*}$ forces can describe the general trend of changes of charge radii along nickel isotopic chain.
Especially the slight odd-even oscillation behavior can be reproduced well against experimental data.
In Ref.~\cite{nickle}, it is mentioned that the trend of changes of charge radii for $^{62-70}\mathrm{Ni}$ isotopes are overestimated by Fayans EDF model.
The sophisticated Fayans energy density functional (EDF) can reproduce the staggering effects of charge radii for Ca isotopes~\cite{Reinhard2017}.
Furthermore, the inverted parabolic-like behavior between $N=20$ and $N=28$ can also be reproduced well.
As demonstrated in Ref.~\cite{nickle}, the isovector component is excluded in pairing interactions.
This may lead to the lack of an isovector component in its pairing part.
In Ref.~\cite{PhysRevC.105.L021303}, the charge radii of even-even Ca-Zn nuclei are obtained by the $ab$ $initio$ coupled cluster theory and nuclear density functional theory, in which a universal pattern from neutron numbers $N=28$ to $N=40$ is exhibited, namely the change in charge radii shows weak dependence on the atomic number.
Actually, our results are in accord with those.

The pioneering work shows that the charge radii formula without considering the unpaired nucleons correlation gives the overestimated OES of charge radii in copper isotopes~\cite{An064101}.
As shown in Fig.~\ref{fig2}~(i), the improved formula can reduce the overestimated odd-even oscillations of charge radii.
In Ref.~\cite{deGroote}, it is found that the charge radii along copper isotopic chain are also slightly overestimated by Fayans DFT model.
Especially close to neutron-rich region, these overestimated OES behaviors are evidently represented.
Therefore, the further performance of optimization-based machine learning method is addressed through Fayans EDF model~\cite{Bollapragada_2021}.

\subsection{Charge radii of $N=28$, $32$, and $34$ isotones}
The emergence of rapidly increasing charge radii is commonly observed across the traditional neutron magic numbers $N=28$, $50$, $82$, and $126$ in the nuclear chart~\cite{KREIM201497,Gorges2019,PhysRevLett.110.032503,PhysRevC.86.034329}.
These phenomena arise from the relative stable binding properties with respect to the adjacent nuclei.
Meanwhile, the new magicity of $N=32$ and $34$ has also been identified by multi-aspects.
Recent studies show that the isotopes with neutron number $N=32$ are proposed to exhibit localized magic behavior due to an observed sudden decrease in binding energy beyond $N=32$, such as in $^{51}$K~($Z=19$)~\cite{PhysRevLett.114.202501}, $^{52}$Ca ($Z=20$)~\cite{RongAn:114101,nature498,PhysRevLett.109.032506}, $^{53}$Sc ($Z=21$)~\cite{PhysRevC.99.064303}, and $^{54}$Ti ($Z=22$)~\cite{JANSSENS200255,PhysRevLett.120.062503}.
This new magicity is also identified from the high excitation energy of the first excited state~\cite{PhysRevLett.114.252501,PhysRevC.31.2226,PhysRevC.70.064303,PRISCIANDARO200117} and the reduced $B(E2)$ transition probabilities~\cite{PhysRevC.100.054317,PhysRevC.84.034318}.
Especially for $^{52}$Ca, the doubly magic nature is also confirmed by two-proton knockout reaction~\cite{PhysRevC.74.021302}.
As demonstrated in Ref.~\cite{PhysRevC.98.024310}, the signature of $N=32$ sub-shell closure is not identified across the vanadium isotopic chain ($Z=23$) and higher proton numbers.

\begin{figure*}[htbp]
\includegraphics[scale=0.32]{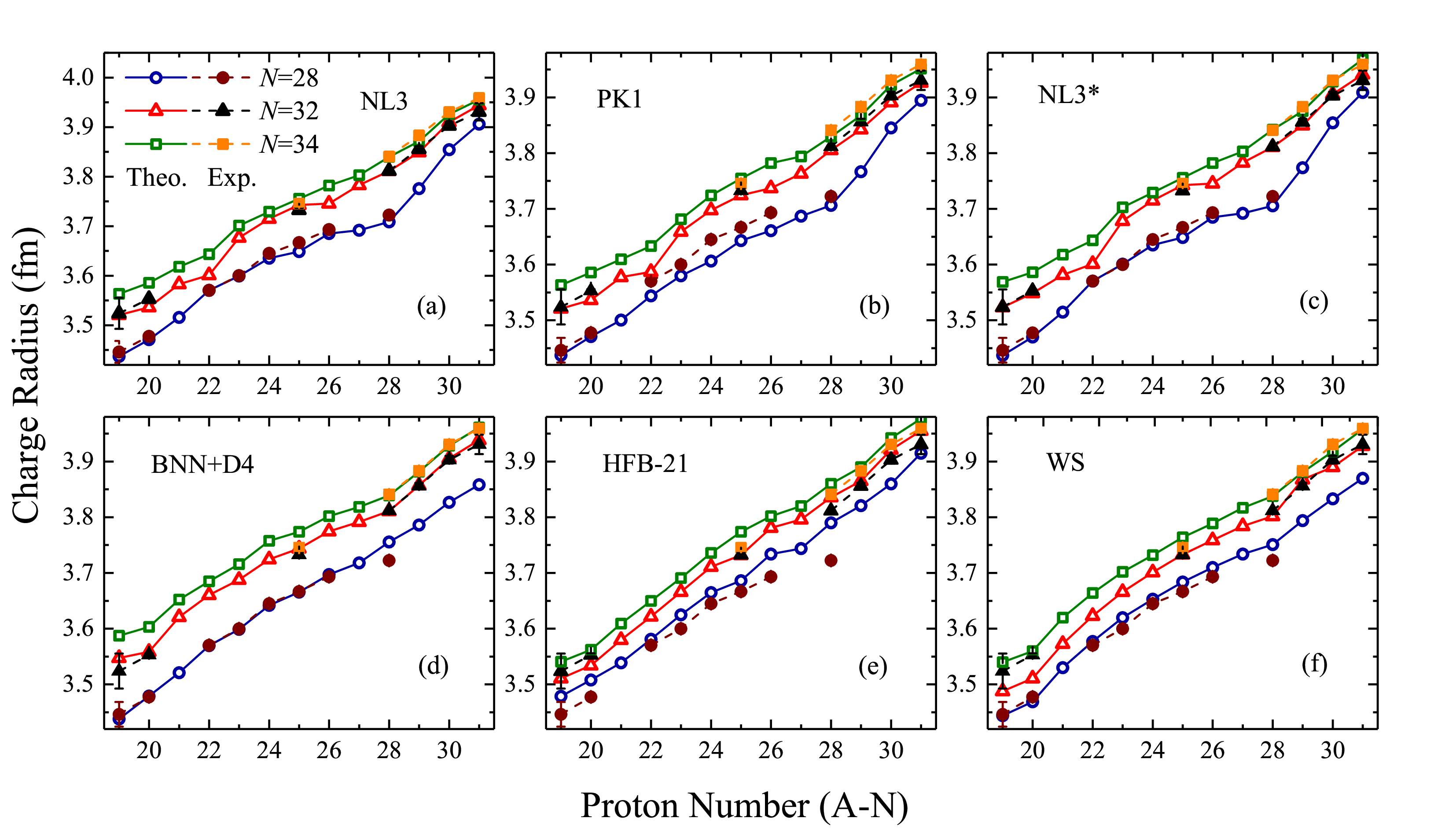}
\caption{Charge radii of $N=28$ (open circle), $32$ (open triangle) and $34$ (open square) isotones as a function of proton number are plotted within (a) NL3, (b) PK1, and (c) NL3$^{*}$ forces.  The results obtained by the (d) BNN+D4~\cite{dong2021novel}, 
(e) HFB-21~\cite{PhysRevC.82.035804}, and (f) WS~\cite{PhysRevC.88.011301} models are also shown for comparison. The corresponding experimental data are taken from Refs.~\cite{ANGELI201369,LI2021101440} (solid marks). } \label{fig3}
\end{figure*}
The closed shell phenomenon has been observed for isotopes with neutron number $N=34$ as well.
The evidence of a $N=34$ sub-shell closure has been suggested in mass of $^{54}$Ca~\cite{PhysRevLett.121.022506} and the $2^{+}$ excitation energy of $^{54}$Ca~\cite{nature502,PhysRevLett.123.142501}.
The same scenario can also be found in $^{52}$Ar isotope~\cite{PhysRevLett.122.072502}.
In Ref.~\cite{PhysRevLett.126.042501}, it does not support the existence of a closed neutron shell in $^{55}$Sc at $N=34$.
However, as suggested in Refs.~\cite{PhysRevC.82.014311,PhysRevC.96.064310,PhysRevC.101.052801} that the $N=34$ sub-shell is weakened at the scandium isotopic chain.
Therefore, further experimental data are urgently required to identify these features.
Close to titanium isotopic chain, the signature of $N=34$ shell closure cannot be observed~\cite{PhysRevC.70.064303,PhysRevC.71.041302}.
This is consistent with Ref.~\cite{PhysRevC.101.052801} where the shell closure effect of $N=34$ vanishes at titanium.
In the Ti and Cr isotopes, the systematics of E(2$^{+}_{1}$)~\cite{PhysRevC.88.024326,PhysRevC.74.064315} and $B(E2)$~\cite{PhysRevC.71.041302,BURGER200529} show no local maximum or minimum at $N=34$.
Latest measurement with the high-precision multireflection time-of-flight technique suggests that the existence of the $N=34$ empirical two-neutron shell gaps for Ti and V cannot be found as well~\cite{PhysRevLett.130.012501}.

In Fig.~\ref{fig3}, the charge radii of nuclei along $N=28$, $32$, and $34$ isotonic chains are plotted.
It clearly shows that the charge radii monotonically increase with the increasing proton numbers.
As the slope of these curves involving different proton numbers is much changed, i.e., the local variations can be found obviously.
The charge radii almost changes linearly from K ($Z=19$) to Fe ($Z=26$) along $N=28$ isotonic chain.
The shrinking trend occurs until nickel ($Z=28$), and then the abrupt change occurs across $Z=28$.
This result suggests that the shell closure effect of charge radii can also be found along $N=28$ isotones.
These calculations for the rms charge radii reveal the characteristic kink at the $Z = 28$ shell closure in accordance with the corresponding experimental radii.

Along $N=32$ isotonic chain, the abrupt changes of charge radii across nuclei with $Z=22$ are observed within NL3 [Fig.~\ref{fig3} (a)], PK1 [Fig.~\ref{fig3}(b)], and NL3$^{*}$ [Fig.~\ref{fig3}(c)] effective forces.
From V to Ga isotopes, the charge radii are changed linearly with the increasing proton numbers, but the results of $^{58}$Fe are slightly deviated.
This can be seen from the Fig.~\ref{fig2} that charge radii of Fe isotopes are systematically underestimated by these three forces.
From K to Ti isotopes, the increasing trend of charge radii are almost linear.
Same scenario can also be found along $N=34$ isotonic chain, but the slope of changes of charge radii from $Z=22$ to $Z=23$ is smaller.
As discussed above, magicity means the nuclei process relatively stable properties.
This pronounced feature can also be remarkably reflected by nuclear size.
As suggested in Ref.~\cite{PhysRevC.98.024310}, the signature of $N=32$ sub-shell closure is not identified across $Z=23$ and higher proton numbers.
This is consistent with our results where the charge radii keep linear increase after the abrupt increase at $Z=22$.
This may provide a signature to verify the new magicity of nuclei with $N=32$.
The same scenario can also be observed along $N=34$ isotonic chain, but the slope of changes of charge radii is smaller.
Consequently, more accurate experimental data are urgently needed in proceeding future.

The latest experimental data show that there has been no sharp change in charge radii across $N=32$~\cite{Koszorus2020mgn}.
Although the magicity of nuclei with $N=32$ has been performed, the shrinking phenomena are not observed along potassium~\cite{PhysRevC.100.034304} and calcium~\cite{Ruiz2016} isotopes.
A more complete picture may shed light on the unexplained phenomena and reveal details of the proton-neutron interaction under the influence of the $N=32$ and $34$ sub-shell closure.
The $N=34$ sub-shell closure has been so far performed in Ca~\cite{PhysRevLett.121.022506,nature502,PhysRevLett.123.142501} and Ar~\cite{PhysRevLett.122.072502} isotopes.
However, the slightly sharp transition in charge radii is also observed across $Z=22$.
A quantitative theoretical interpretation of the well-known sub-shell effect in $N$ has been hampered by the fact that detailed calculations use somewhat inconsistent models in treating even-$N$ and odd-$N$ nuclei.
The present case of a pronounced $Z$ dependence of the charge radii along $N=28$, $32$, and $34$ isotonic chains may provide the clue for a clear-cut phenomenological interpretation.

\begin{figure}[htbp]
\includegraphics[scale=0.36]{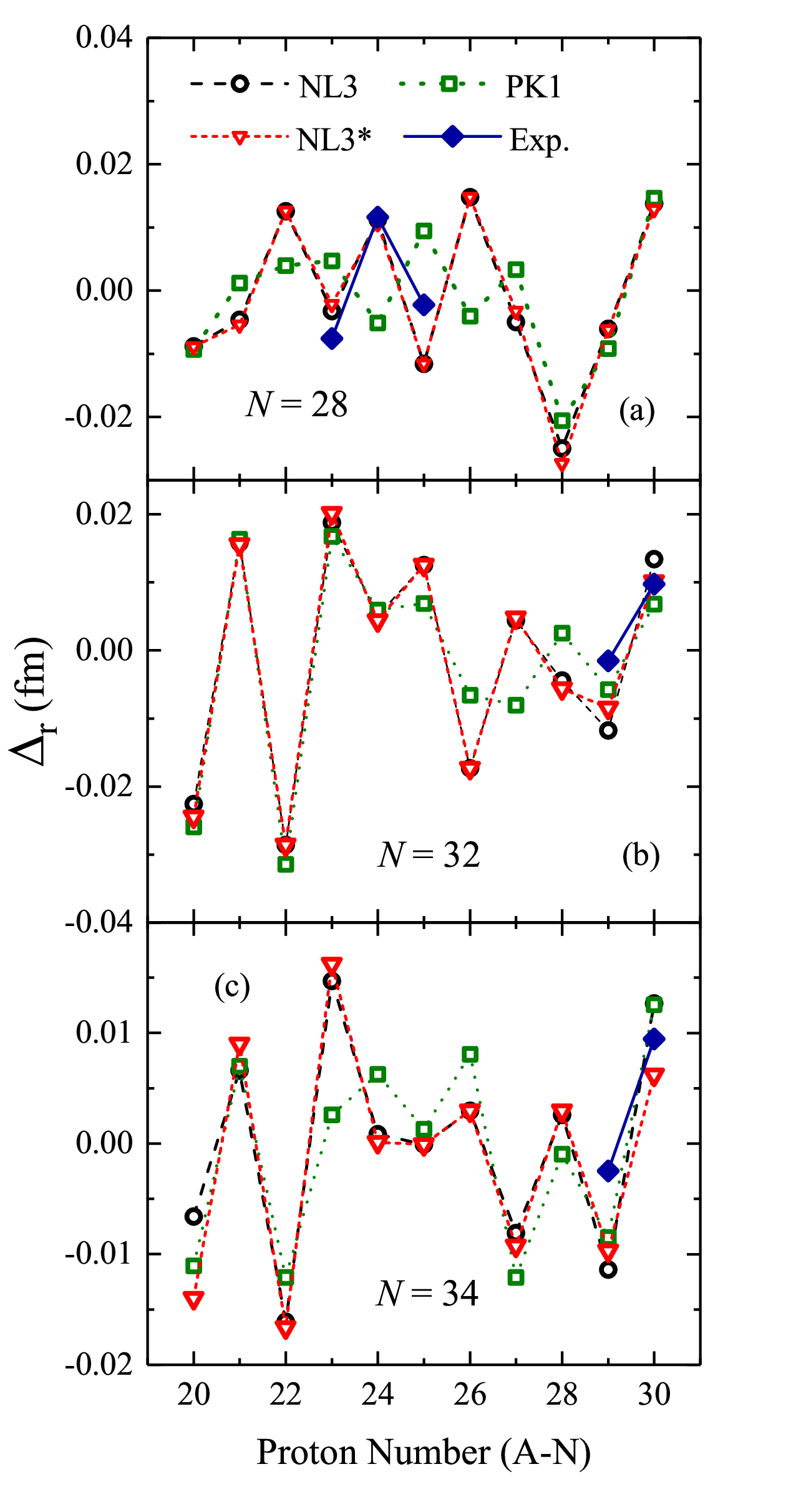}
\caption{Odd-even staggering of charge radii versus (a) $N=28$, (b) $32$, and (c) $34$ isotones as a function of proton number are plotted within NL3 (open diamond), PK1 (open square), and NL3$^{*}$ (solid triangle) forces. The corresponding experimental data are taken from Refs.~\cite{ANGELI201369,LI2021101440}.} \label{fig4}
\end{figure}
Such a comparison should be helpful in providing an accessible understanding of nuclear charge radii and in making reliable predictions for those nuclei lacking an experimental value.
Although the systematic evolution of the charge radii over a large range of proton numbers seems to be consistent with the presence of a noticeable sub-shell effects at $N = 32$ and $34$, it misses the pronounced experimental data to identify these results.
In order to further examine the universal trend of nuclear charge radii, it will be also instructive to gather similar data on the isotonic chains at neutron number $N=28$, $32$, and $34$.
In Figs.~\ref{fig3}(d), \ref{fig3}(e), and \ref{fig3}(f), the results obtained by Bayesian neural network (BNN+D4)~\cite{dong2021novel}, HFB-21~\cite{PhysRevC.82.035804}, and WS~\cite{PhysRevC.88.011301} are also shown for comparison.

In Fig.~\ref{fig3} (d), the change of charge radii obtained by BNN+D4 is almost linear along $N=28$ isotonic chain.
The abrupt increases are predicted along $N=32$ and $34$ isotonic chains at the proton numbers $Z=20$ and $28$, respectively.
This can be understood clearly where the pronounced shell effects are emphasized in the calibrated protocol~\cite{dong2021novel}.
For HFB-21 model~\cite{PhysRevC.82.035804}, the charge radii of $N=28$, $32$, and $34$ isotonic chains change linearly with the increasing proton numbers.
While the slight shrinking occurs at $Z=27$, this may be a result from the unpaired proton.
This means the strong shell effect is weakened in Skyrme functionals.
As demonstrated in Refs.~\cite{TAJIMA1993434,Gorges2019}, the calculations in non-relativistic density functional theories (DFT) based on conventional functionals are unable to reproduce the kink well at magic numbers.
Such as Skyrme functionals utilizing SV-min force generally cannot reproduce the characteristic kink typically found at shell closure along Ni isotopic chain~\cite{nickle}.
As demonstrated in Ref.~\cite{PhysRevLett.74.3744}, shell closure effect results from the rather small isospin dependence of spin-orbit interactions.
Meanwhile, the great deviations are also observed in charge radii at the proton numbers $Z=26$ and $28$.
For WS model, the shell effect and isospin effect in nuclear charge radii are systematically investigated~\cite{PhysRevC.88.011301}.
Along $N=32$ and $34$ isotonic chains, the $Z=20$ and $28$ shell closure are observed apparently.
Furthermore, the rapid increase of charge radii can also be found across $Z=20$, but weakened in our calculations.
And the charge radii of $^{51}$K and $^{52}$Ca isotopes are underestimated in the WS model.
From $Z=29$-$31$ isotopic chains, charge radii of $N=32$ isotones can be reproduced well by NL3, PK1, NL3$^{*}$, BNN+D4, and WS models.
But the results obtained by HFB-21 are slightly overestimated against the experimental data.

Considering the neutron-proton correlation in describing the evolution of nuclear charge radii, the increasing trends of charge radii along $N=32$ and $34$ isotonic chains seem to suggest the limitation of new magicity at the proton number $Z=22$. Meanwhile, it leads to the vanished magicity of $Z=28$ along the neutron numbers $N=32$ and $N=34$ isotones. It is also mentioned that the magicity at $Z=28$ is weakened along $N=32$ and $34$ isotonic chains for the HFB-21 model, but the BNN+D4 and WS models show the magicity at $Z=28$ as well as at $Z=20$.

In general, the odd-even staggering in charge radii describes the fact that the nuclear charge radii of odd-$N$ isotopes are smaller than the averages of their even-$N$ neighbours.
In order to emphasize these local variations, a three-point formula has been employed to extract the odd-even oscillation behaviors of nuclear charge radii along a specific isotopic chain~\cite{Reinhard2017,An:2020qgp}.
As shown in Fig.~\ref{fig3}, the abrupt changes of charge radii across $Z=22$ obtained by NL3, PK1, and NL3$^{*}$ effective forces are also shown along $N=32$ and $34$ isotones, but the latter with a less slope.
For further inspecting this sudden change in nuclear charge radii, the OES of charge radii along $N=28$, $32$, and $34$ isotones are also calculated by this three-point formula.
The definition is rewritten as $\Delta_{r}(Z,N)=\frac{1}{2}[2R(Z,N)-R(Z-1,N)-R(Z+1,N)]$,
where $R(Z,N)$ is rms charge radius along isotonic chain.

As shown in Fig.~\ref{fig4}, the OES of charge radii of $N=28$, $32$, and $34$ isotones as a function of proton number are plotted within NL3, PK1, and NL3$^{*}$ effective forces.
For the $N=28$ isotones,the results obtained by NL3 and NL3$^{*}$ forces agree well against the experimental data along $Z=23$-$25$ isotopes, while PK1 force give the opposite sign until $Z=26$.
At $Z=20$ and $28$, these three methods gives the similar trend, in which the OES of charge radii presents the anomalous behavior.
As demonstrated in Ref.~\cite{PhysRevC.105.014325}, this results from the strong shell closure effect.

Along $N=32$ and $34$ isotones, NL3, PK1, and NL3$^{*}$ effective forces can reproduce the experimental data from $Z=29$-$30$.
Both the NL3 and NL3$^{*}$ forces give the almost identical results, while PK1 gives the opposite sign at $Z=27$ along $N=32$ isotone.
The same scenarios can also be found at $Z=24$ along $N=34$ isotone.
Here, we greatly pay more attention to the local variations of charge radii across the proton number $Z=22$.
For $N=32$ and $34$ isotones, one can find that the amplitudes of OES in charge radii are increased at $Z=22$.
This is consistent with the trend of changes of charge radii as shown in Fig.~\ref{fig3}.
In addition, one can find that the amplitudes of OES in charge radii are comparative for $Z=20$ and $Z=22$ isotones.

\section{Conclusions}\label{sec3}
The neutron-proton ($np$) correlation originating from the unpaired neutron and proton are taken into account in describing nuclear charge radii.
The systematic study of charge radii for nuclei with proton numbers $Z=19$-$29$ isotopes are performed.
The odd-even staggering (OES) and the inverted parabolic-like behaviors in charge radii are reproduced well along potassium and calcium isotopes.
Although the NL3, PK1, and NL3$^{*}$ effective forces give the similar trend, the PK1 parameter set systematically underestimates the results.
Beyond $N=28$ shell closure, the sharp increases are reproduced remarkably well.
Especially the charge radii of nickel and copper isotopes can be reproduced against the latest experimental data well.
It can be found that the increasing trend of charge radii beyond $N=28$ shell closure is almost independent of proton number, which is in accord with Ref.~\cite{PhysRevC.105.L021303}.
Our calculations can give the shrunk phenomenon of charge radii of Sc isotopes at $N=20$, but the trend is underestimated.

In general, sudden drop in nucleon separation energies, a larger $2^{+}_{1}$ excitation energy, and a smaller transition probability as compared to the neighbor isotones (or isotopes) are commonly considered as manifestations of the magic character of a nucleus.
Along isotopic chains, unexpected increases of charge radii are not observed across neutron number $N=32$~\cite{PhysRevC.100.034304,Koszorus2020mgn}.
The shell effect leads to the kink in the isotopic chains, where charge radii vary smoothly closing to neutron magic numbers, and then the sudden increase occurs after crossing filled shells.
In our calculations, the calculated results monotonically increase with the increasing neutron number along $Z=19$-$29$ isotopic chains.
Therefore, it is difficult to identify the new magicity of nuclei with neutron number $N=32$ and $34$ along a long isotopic chain.

However, a high interest is the local variations of charge radii along isotonic chains.
In our calculations, the abrupt change of nuclear charge radii across $Z=28$ is evidently shown along $N=28$ isotonic chain, but this could be weakened across $Z=20$.
Along $N=32$ and $34$ isotones, the abrupt increases of charge radii are also predicted across the proton number $Z=22$, but the latter with a lower slope.
In potassium and calcium isotopes, $N=32$ possesses the magicity in the view of relative stable binding properties.
However, the shrinking phenomena cannot be observed in charge radii of nuclei with featuring $N=32$.
Our results suggest that the trend of changes in charge radii is almost linear from potassium $(Z=19)$ to titanium $(Z=22)$.
This is consistent with the Ref.~\cite{PhysRevC.98.024310} where the signature of $N=32$ sub-shell closure is not identified across V ($Z=23$) and nucleus with higher proton numbers.
The same scenario can also be observed along $N=34$ isotonic chain, but more available results are needed.
An indication of the robustness of a new shell closure may be identified from these aspects.

An available description of the nuclear size can provide access to understand new physics beyond standard model~\cite{RevModPhys.90.025008} and serve as input quantities in astrophysical study~\cite{ARNOULD2020103766}.
As demonstrated in Ref.~\cite{Perera:2023hnv}, proton-neutron interaction is responsible for the fine structure of charge radii.
In our calculations, the unpaired neutron-proton correlation is tentatively incorporated into the root-mean-square charge radii formula.
The overestimated odd-even staggering in charge radii are definitely improved by this modified formula.
Although this modified formula can reproduce the local variations of nuclear charge radii, the self-consistently microscopic form is urgently required.
It is known that the linear relationship between the difference of the charge radii of mirror nuclei and nuclear symmetry energy can be used to inspect the isovector forces~\cite{PhysRevLett.130.032501,PhysRevC.97.014314}.
Therefore, more available models are needed in describing nuclear size and more data about the charge radii of mirror partner nuclei are required in experiments.
\section*{Acknowledgments}
This work is supported in part by the National Key R$\&$D Program of China under Grant No. 2023YFA1606401 and the National Natural Science Foundation of China under Grants No. 12047513, No. 12135004, No. 11635003, No. 11961141004, No. 11025524, No. 11161130520, the National Basic Research Program of China under Grant No. 2010CB832903. X. J. is grateful for the support of the National Natural Science Foundation of China under Grants No. 11705118. L.-G. C. is grateful for the support of the National Natural Science Foundation of China under Grants No. 12275025, No. 11975096 and the Fundamental Research Funds for the Central Universities (2020NTST06).

\bibliography{refsnew}

\end{document}